# Improving operational flexibility of integrated energy system with uncertain renewable generations considering thermal inertia of buildings

Yang Li [a,*], Chunling Wang [a], Guoqing Li [a], Jinlong Wang [b], Dongbo Zhao [c], Chen Chen [d]

[a] School of Electrical Engineering, Northeast Electric Power University, Jilin 132012, China

[b] State Grid Zhucheng Power Supply Company, Zhucheng 262200, China

[c] Energy Systems Division, Argonne National Laboratory, Lemont 60439, USA

[d] School of Electrical Engineering, Xi'an Jiaotong University, Xi'an 710049, China

**Abstract**: Insufficient flexibility in system operation caused by traditional "heat-set" operating modes of combined heat and power (CHP) units in winter heating periods is a key issue that limits renewable energy consumption. In order to reduce the curtailment of renewable energy resources through improving the operational flexibility, a novel optimal scheduling model based on chance-constrained programming (CCP), aiming at minimizing the lowest generation cost, is proposed for a small-scale integrated energy system (IES) with CHP units, thermal power units, renewable generations and representative auxiliary equipments. In this model, due to the uncertainties of renewable generations including wind turbines and photovoltaic units, the probabilistic spinning reserves are supplied in the form of chance-constrained; from the perspective of user experience, a heating load model is built with consideration of heat comfort and inertia in buildings. To solve the model, a solution approach based on sequence operation theory (SOT) is developed, where the original CCP-based scheduling model is tackled into a solvable mixed-integer linear programming (MILP) formulation by converting a chance constraint into its deterministic equivalence class, and thereby is solved via the CPLEX solver. The simulation results on the modified IEEE 30-bus system demonstrate that the presented method manages to improve operational flexibility of the IES with uncertain renewable generations by comprehensively leveraging thermal inertia of buildings and different kinds of auxiliary equipments, which provides a fundamental way for promoting renewable energy consumption.

## NOMENCLATURE

**Acronyms**

| | |
|---|---|
| IES | Integrated energy system |
| CCP | Chance-constrained programming |
| CHP | Combined heat and power |
| BESS | Battery Energy storage system |
| EB | Electric boiler |
| HST | Heat storage tank |
| DG | Distributed generation |
| SOT | Sequence operation theory |
| MILP | Mixed-integer linear programming |
| MG | Microgrid |
| PV | Photovoltaic |
| WT | Wind turbine |
| Zn-Br | Zinc-bromine |
| PDF | Probability density function |

**Symbols**

| | |
|---|---|
| $t$ | Time period (h) |
| $v$ | Actual wind speed (m/s) |
| $F$ | Surface area of the building (m$^2$) |
| $V$ | Building volume (m$^3$) |
| $S_{BESS}$ | Capacity of the BESS (MWh) |
| $C_{HST}$ | Capacity of the HST (MWh) |
| $P_{h,i}$ | Heating power of CHP unit $i$ (MW) |
| $P_{e,i}$ | Electric power of the CHP unit $i$ (MW) |
| $P_{ZS,i}$ | Electric power under condensing condition of CHP unit $i$ (MW) |
| $P_{EB}$ | EB electric power (MW) |
| $P^{CH}$ | BESS charge power (MW) |
| $P^{DH}$ | BESS discharge power (MW) |
| $P^{PV}$ | PV output power (MW) |
| $R^{BESS}$ | Spinning reserve of the BESS (MW) |
| $R_i$ | Spinning reserve of thermal power unit $i$ (MW) |
| $R_{e,i}$ | Spinning reserve of CHP unit $i$ (MW) |
| $p_s$ | Rated outputs of WT (MW) |
| $\rho_{air}$ | Density of indoor air (kg/m$^3$) |
| $A_{pv}$ | Radiation area of the PV (m$^2$) |
| $T_{id}$ | Indoor temperature (°C) |
| $T_{od}$ | Outdoor temperature (°C) |
| $\eta_{EB}$ | EB electric-to-heat conversion efficiency (%) |
| $\gamma_{CH}$ | BESS charge efficiency (%) |
| $\gamma_{DH}$ | BESS discharge efficiency (%) |
| $\omega_i$ | Spinning reserve cost of thermal power unit $i$ ($/MWh) |
| $\delta_i$ | Spinning reserve cost of CHP unit $i$ ($/MWh) |
| $\mu$ | Spinning reserve cost of BESS ($/MWh) |

**Subscripts**

| | |
|---|---|
| $in$ | Cut-in |
| $out$ | Cut-out |
| $min$ | Minimum value |
| $max$ | Maximum value |
| $a$ | Probabilistic sequence of WT outputs |
| $b$ | Probabilistic sequence of PV outputs |
| $c$ | Probabilistic sequence of joint outputs |
| $s$ | Rated value |
| $n$ | Number of thermal power unit |
| $r$ | Solar irradiance |
| $N$ | Number of CHP |

**Superscripts**

| | |
|---|---|
| $W$ | Wind |
| $CH$ | Charge |
| $DC$ | Discharge |

## 1 Introduction

With increasingly serious energy crisis and environmental pollution, use of renewable energies has been regarded as one of

---

the most important choices for ensuring secure and sustainable energy supply [1], due to the advantages of renewable energies such as cleanness, easy availability, low cost, and abundance [2]. Unfortunately, the increasing uncertainties of renewable generations will pose huge challenges in the operation of today's power systems [3]. Furthermore, the traditional 'heat-set' constraints and growing curtailment of renewable energy resources greatly limit the flexibility and economy of the system operation [4]. At the same time, the presence of an integrated energy system (IES), which comprehensively utilizes multiple energies in a region to achieve coordinated planning and operations among multiple energy forms [5], provides more regulatory means available for enabling greater consumption of renewable energies [6]. By utilizing the time and space complementary characteristics of energy and power of multiple power sources such as wind turbines (WT) and photovoltaic (PV) units, coordinated scheduling of a multi-energy system can improve the operational flexibility of the system and expand the space for renewable energy consumption, thereby providing a new way to ensure secure and sustainable energy supplies. Therefore, how to improve the flexibility of IES to promote renewable energy consumption is of great significance [7].

### 1.1 Literature review

Up to now, some pioneering works regarding IES have been performed for promoting renewable energy consumption. (1) In terms of energy storage, a heat storage tank (HST) and an electric boiler (EB) are configured to improve the flexibility of combined heat and power (CHP) units and reduce wind power curtailment in [8]. Reference [9] proposes a modeling and optimization approach for the district heating system with CHP units and battery energy storage system (BESS). References [10] propose a measure for supplying heating by using EBs and utilizes pumped hydro for energy storage, resulting in reducing the curtailment of wind power. Reference [11] evaluates the impact of EBs and heat pumps on the improvement of system flexibility. In [12], a strategy is proposed for renewable integration by using the flexibility of electric vehicles. Reference [13] presents an integrated dispatch approach for CHP plant with phase-change HST with consideration of heat transfer process. Reference [5] presents an optimal operation model of IES which combines the thermal inertia of the central heating network with the thermal inertia of the building to improve the consumption of wind power. The above works consider the complementarity between power systems and heating systems, and enhance the energy optimization and allocation capability in a wider space-time range. However, there is little discussion of the consumption of wind energy curtailment in the joint operation of the BESS, HSTs and EBs in existing works. (2) Regarding thermal comfort, reference [14] proposes a demand response management and thermal comfort optimization control algorithm for microgrids (MGs) with renewable energy and energy storage units. Reference [15] proposes a bi-level optimization model of IES considering the thermal comfort of heat users, thus determines heating load demands according to indoor temperature. (3) Regarding distributed generations (DGs), an optimal energy management method for a grid-connected PV battery hybrid system to utilize solar energy and benefit users from the demand side is proposed in [16]. (4) Regarding IES scheduling model, reference [6] presents a novel optimal operation model of IES with consideration of thermal inertia of a district heating network and buildings to promote wind power consumption, and reference [17] models integrated community energy system for performing optimal planning and operation of the system by simulating the local exchange and community level operation. These approaches prove effective in promoting renewable energy consumptions through coordinated scheduling of multiple energy forms. Unfortunately, these references do not simultaneously consider the renewable uncertainties and thermal inertia of buildings, which limits the system operational flexibility to a certain extent. Moreover, most of the previous studies haven't taken into account the spinning reserve provision caused by prediction errors between the DG predicted outputs and the actual outputs.

Due to the intermittency and uncertainty of DG outputs, the system needs more flexibility for the consumption of renewable energies, especially during the winter heating period [18]. However, traditional power sources are mainly made up of coal-fired generators with limited adjustment capability [19], and the "heat-set" mode reduces the adjustment capability and compresses the space for the accommodation of renewable energies [20]. Therefore, how to improve the operational flexibility of the system through thermal-electric decoupling is an urgent and challenging problem restricting renewable energy consumption.

Although basic optimal scheduling problems of IES have been solved, there are still the following research gaps in this field. (1) The uncertainty of renewable energies has an important impact on a scheduling strategy [21], but at the same time, handling the uncertainty of multiple kinds of renewable generations has always been a difficult problem. (2) In addition, only several previous works take into account the heating load model satisfying the thermal comfort of heat customers. (3) Most importantly, to the authors' best knowledge, until now there has been little research reported on IES scheduling that improves the system operational flexibility by comprehensively leveraging different kinds of auxiliary equipments as thermal-electric decoupling measures to address challenges associated with renewable curtailments.

### 1.2 Contributions of this work

In order to address the above-mentioned issues, a novel optimal scheduling model based on chance-constrained programming (CCP), aiming at minimizing the lowest generation cost, is proposed for a small-scale IES. The main contributions of this work include the following three folds:

(1) To reduce renewable curtailment through improving the operational flexibility, a chance-constrained programming-based scheduling model is proposed for a small-scale IES with uncertain renewable generations and different kinds of auxiliary equipments by considering thermal inertia of buildings. In this model, due to the uncertainty of multiple renewable generations, the spinning reserves are provided in the form of chance constraints; and the heating load model described in this paper is depicted via thermal comfort requirements of end users and the transient heat balance equation of the buildings.

(2) A sequence operation theory (SOT)-based solution approach is developed for solving the above scheduling model. Through converting a chance constraint into its deterministic equivalence class with the use of SOT, the original scheduling model is formulated as a mixed-integer linear programming (MILP), and thereby solved by utilizing the CPLEX solver, which is a new general solution methodology to address CCP problems.
(3) The simulation results on the modified IEEE 30-bus system demonstrate that the proposed method manages to improve the operational flexibility of the IES by comprehensively utilizing multiple auxiliary equipments as thermal-electric decoupling measures, and that the IES operation is able to strike a balance between economy and reliability through choosing a proper confidence level of probabilistic spinning reserve constraints. In addition, consideration of the thermal inertia of buildings proves to be beneficial to improve the system operational flexibility and thereby promote renewable energy consumption. Besides, due to the different charging and discharging periods of various auxiliary equipments, there is a certain energy reserve in each period during system operation, which is beneficial for strengthening the reliability of the energy supply.

## 1.3 Organizational arrangement

The rest of this work is structured as follows. The structure modeling of an IES is provided in Section 2. The proposed IES scheduling model and its solution approach are respectively given in Section 3 and 4. Section 5 investigates case studies, and finally conclusions are made in Section 6.

## 2 Structure modeling of integrated energy system

In order to fully demonstrate the physical model of the IES, this section first describes the overall framework of the IES and then models each unit separately.

### 2.1 Overall framework of the system

The overall structure of a small-scale IES is shown in Fig. 1. The system integrates two forms of energy: heat and electricity. And furthermore, a wind turbine, a photovoltaic unit, four thermal power units, two CHP units and representative auxiliary equipments such as BESS, HST and HB are integrated into the IES. The HB, CHP units and the building on the load side are three electro-thermal coupling units of the system. The HST directly increases the flexibility of the CHP unit; the EB is able to absorb part of the DG outputs; the BESS, as an energy buffer, is able to shift electrical loads in time by continuously adjusting its working modes according to the change of DG outputs.

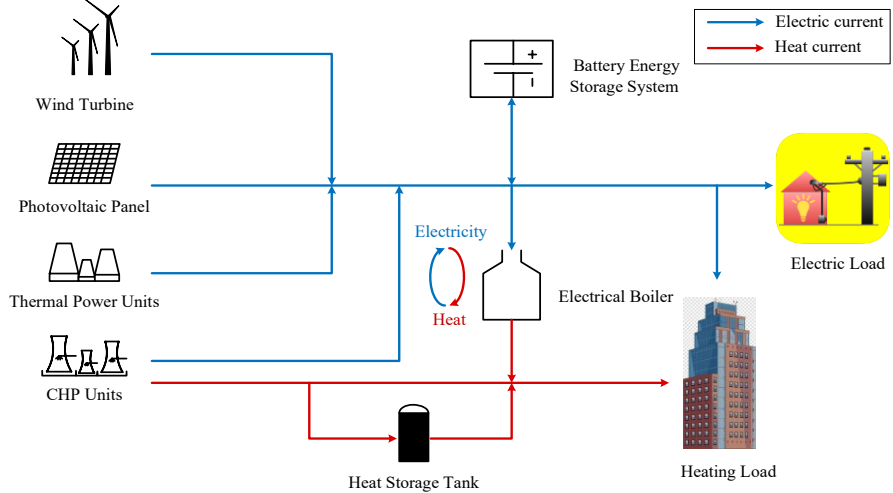

**Fig. 1.** Overall structure of the IES

### 2.2 Combined heat and power unit model

There are two main types according to the steam turbines used for CHP [22]. In this paper, the most common extraction steam CHP unit is chosen for example. The area surrounded by the closed black curve ABCD represents the traditional electro-thermal operation characteristics in Fig. 2. According to the operating principle of the extraction unit, the output model is as follows [23]:

$$P_{ZS,it} = P_{e,it} + C_{V,i} P_{h,it} \tag{1}$$

where $C_{V,i}$ is the thermal power ratio; $P_{h,it}$ is the heating power; $P_{e,it}$ is the electric power; $P_{ZS,it}$ is the electric power under condensing condition.

After a traditional CHP unit is installed with HST, its electro-thermal operation characteristics will change, as shown in the area surrounded by the closed red curve AEFGHI in Fig. 2 [8]. The heating range of the unit is expanded from $[0, P_{h,\max}]$ to $[0, P_{hc,\max}]$, the electric power range is expanded from $[P_{e,K}, P_{e,J}]$ to $[P_{e,M}, P_{e,L}]$ at the same heating power $P_{h,it}$, The HST significantly increases the adjustability of the unit and reduces the electro-thermal coupling characteristics.

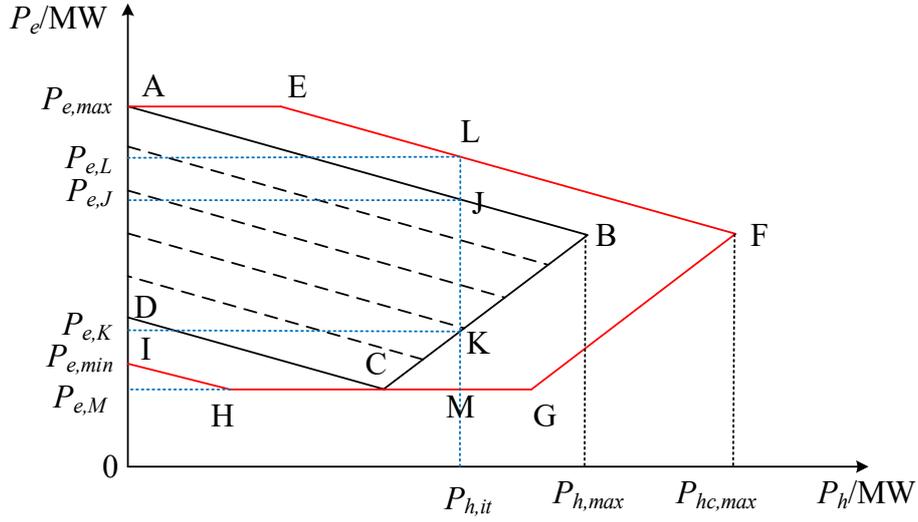

Fig. 2. Electro-thermal characteristic curves of CHP units

### 2.3 Electric boiler model

As a coupling unit, an electric boiler plays a double peak regulation role by converting electrical energy into heat for heating at night. For one thing, this reduces the heating load and the minimum power outputs of the CHP units, which improves the system operational flexibility; for another, the total electric demand on the load side is increased during the load valley period, which further expands the adjustment range of the power output of the CHP units. In addition, the EB heating efficiency can reach 95% with no pollutant emissions such as $CO_2$, $SO_2$. The output model of the EB is given as follows [15]:

$$P_{EB,t}^r = P_{EB,t} \eta_{EB} \tag{2}$$

where $P_{EB,t}$ and $P_{EB,t}^r$ are the electric power and heating power of EB in period $t$, respectively; $\eta_{EB}$ is the electric-to-heat conversion efficiency.

### 2.4 Battery energy storage system model

The BESS can flexibly adjust the output and reasonably transfer electric energy, and thereby smoothing the fluctuations of the DGs outputs. During the off-peak period, the BESS is charged by the surplus power; while during the peak-load period, the stored energy in the BESS is discharged [9]. There are many types of batteries that can be used as energy buffers in IES. Compared with other flow batteries such as sodium-sulfur battery [24], lithium-ion battery, and lead-acid battery, a zinc-bromine battery (Zn-Br) flow battery has some significant advantages, like relatively higher energy density, longer service life, and lower costs [25], and thereby it has been widely used for renewable energy integration and independent power generation systems. Since this work aims to promote renewable energy consumption, Zn-Br batteries are here selected to build the used BESS. The available capacities of the BESS in period $t$ is [26]

$$S_t = S_{t-1} + (\gamma_{CH} P_t^{CH} - P_t^{DC}/\gamma_{DC})\Delta t \tag{3}$$

where $S_t$ and $S_{t-1}$ are the capacities of the BESS in period $t$ and $t-1$; $P_t^{CH}$ and $\gamma_{CH}$ are the charge power and efficiency, respectively; $P_t^{DC}$ and $\gamma_{DC}$ are respectively the discharge power and efficiency; $\Delta t$ denotes a time interval.

### 2.5 Heat storage tank model

For a CHP unit with a HST, the conventional 'heat-set' operation mode is completely broken, i.e. the power loads are no longer limited by the heating load demands. The CHP can achieve flexible adjustment according to the renewable curtailment condition. When the amount of renewable curtailment is large, the HST performs heating to reduce the CHP unit outputs and thereby promote the consumption of the DGs; while on the contrary, when the amount of renewable curtailment is small, the HST stores heat. The available capacity of the HST is formulated as [13]

$$C_{i,t} = C_{i,t-1} - P_{h,it}^c \Delta t \tag{4}$$

where $C_{i,t}$ is the capacity of HST in period $t$; $P_{h,it}^c$ is the power of heat storage and release in period $t$. Note that $P_{h,it}^c$ is a positive number when the heat is released, while it is negative when the heat is stored.

### 2.6 Probabilistic wind turbine model

It's known that WT outputs are related to the uncertainty of wind speed. Studies have shown that the probability density function (PDF) of the average wind speed can be described as follows [28]:

$$f_V(v) = m/\varepsilon (v/\varepsilon)^{m-1} \exp[-(v/\varepsilon)^m] \qquad (5)$$

where $v$ is the actual wind speed; $\varepsilon$ is the scale factor, which reflects the average wind speed during a certain time period; $m$ is a shape factor that describes the shape of the wind speed distribution function. The relationship between WT outputs and the actual wind speed can be described as [29]

$$P^W = \begin{cases} 0 & v < v_{in}, v \geq v_{out} \\ \dfrac{v - v_{in}}{v_s - v_{in}} p_s & v_{in} \leq v < v_s \\ p_s & v_s \leq v < v_{out} \end{cases} \qquad (6)$$

where $p_s$ is the WT rated power output; $v_s$ is the rated wind speed; $v_{in}$ and $v_{out}$ denote the cut-in and cut-out wind speeds. The cumulative probability distribution of the WT power output is formulated as

$$F(p^W) = \begin{cases} 0, & p^W < 0 \\ 1 - \exp\left\{-\left[\dfrac{(1 + \dfrac{h \cdot p^W}{p_s})v_{in}}{\varepsilon}\right]^m\right\} + \exp[-(v_{out}/\varepsilon)^m], & 0 \leq p^W < p_s \\ 1, & p^W \geq p_s \end{cases} \qquad (7)$$

Where $h = (v_s/v_{in}) - 1$.

### 2.7 Probabilistic photovoltaic model

The PV outputs are mainly affected by solar irradiance, which is related to the location of the sun, the geographical location of PV generation equipment and weather conditions. Previous studies have shown that the solar irradiance within a certain time period is approximated as the Beta distribution [30], which is a continuous distribution on the open interval (0,1). The PDF of solar irradiance $r$ is

$$f_r(r) = \frac{\Gamma(\lambda_1) + \Gamma(\lambda_2)}{\Gamma(\lambda_1)\Gamma(\lambda_2)} \left(\frac{r}{r_{max}}\right)^{\lambda_1 - 1} \left(1 - \frac{r}{r_{max}}\right)^{\lambda_2 - 1} \qquad (8)$$

where $r_{max}$ is the maximum solar irradiance; $\lambda_1$ and $\lambda_2$ are the shape parameters of the Beta distribution; the Gamma function $\Gamma$ is expressed as $\Gamma(\lambda) = \int_0^{+\infty} \rho^{\lambda - 1} e^{-\rho} d\rho$, here $\rho$ is an integer variable. The power output of the PV is [27]

$$P^{PV} = r A_{pv} \eta_{pv} \qquad (9)$$

where $P^{PV}$ is the PV power output, $A_{pv}$ is the PV radiation area, and $\eta_{pv}$ denotes the conversion efficiency.

Based on the relationship between the PV output and the solar irradiance, the PDF of $P^{PV}$ can be obtained as follows [31]:

$$f_p(P^{PV}) = \frac{\Gamma(\lambda_1) + \Gamma(\lambda_2)}{\Gamma(\lambda_1)\Gamma(\lambda_2)} \left(\frac{P^{PV}}{P_{max}^{PV}}\right)^{\lambda_1 - 1} \left(1 - \frac{P^{PV}}{P_{max}^{PV}}\right)^{\lambda_2 - 1} \qquad (10)$$

where $P_{max}^{PV}$ represents the maximum value of $P^{PV}$.

### 2.8 Modeling of heating load

In this study, the thermal capacity of a building with heating systems is taken as the heating load demand. Firstly, the fuzziness of the heat demands and the thermal inertia of a building are analyzed, and then the heating load model is obtained.

#### 2.8.1 Fuzziness of the heat demand

The heating load demands refer to the thermal capacity provided by the heating system that enables the building to maintain a certain indoor temperature. The indoor temperature here needs to meet the human body's requirements for thermal comfort, that is, thermal comfort temperature. According to related researches, thermal comfort temperature usually has a fuzzy range. Based on the theory of fuzzy mathematics, the membership function is used in this section to describe indoor thermal comfort temperature and then participate in the optimization of the heating system. As shown in Fig. 3, the trapezoidal membership function is used to describe the indoor thermal comfort temperature in winter [32]. Here, the temperature range corresponding to the membership degree 1 is taken as the range of the indoor comfort temperature.

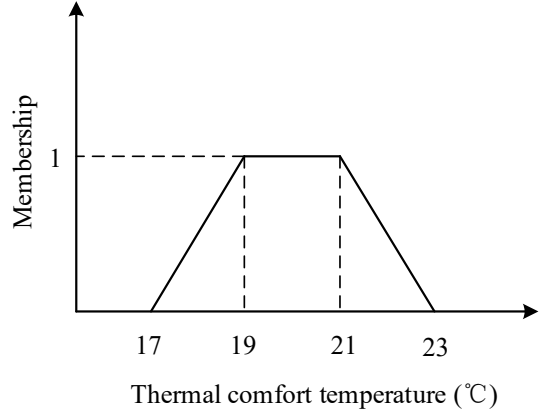

Fig. 3. The trapezoidal membership function

### 2.8.2 Thermal inertia of a building

The envelope structure of a building is able to store a certain amount of heat. For a greater heat storage capacity of a building, the internal temperature of the building is less susceptible to the external environment, and correspondingly, the thermal inertia of the building is greater [33]. According to the theory of heat balance, when the heat supply in the building is greater than the heat dissipation, the indoor temperature will rise due to heat accumulation; on the contrary, when the heat supply in the building is less than the heat dissipation, the indoor temperature will drop. The transient heat balance equation of the building, which is used to describe the effect of heat changes supplied by the heating system on the building temperature, is bulit in this section, and thereby a connection between the heat and temperature is established. The transient heat balance equation of the building can be described as the following first-order differential equation [34]:

$$\frac{dT_{id}}{dt} = \frac{P_{lt}^r - (T_{id}(t) - T_{od}(t)) \cdot K \cdot F}{c_{air} \cdot \rho_{air} \cdot V} \quad (11)$$

where $T_{id}(t)$ and $T_{od}(t)$ are respectively the indoor and outdoor temperatures in period $t$; $K$ is the comprehensive heat transfer coefficient of the building; $F$ is the surface area of the building; $V$ is the building volume; $C_{air}$ is the specific heat capacity of indoor air; $\rho_{air}$ is the density of indoor air; $P_{lt}^r$ is the heating power in period $t$.

Since the outdoor temperature changes very little in a short time, it can be assumed that the outdoor temperature is constant for a certain time period. Thus, according to the general solution of the first-order linear homogeneous differential equation, the linearized heat balance equation can be obtained [34].

$$T_{id}(t) = (-\frac{K \cdot F}{c_{air} \cdot \rho_{air} \cdot V} \cdot \Delta t) \cdot (T_{id}(t-1) - T_{od}(t) - P_{lt}^r \cdot \frac{1}{K \cdot F}) + (T_{od}(t) + P_{lt}^r \cdot \frac{1}{K \cdot F}) \quad (12)$$

And then, the heating load model can be obtained by transforming Eq. (12), which is given by

$$P_{lt}^r = \frac{[T_{id}(t) - T_{od}(t)] + \frac{K \cdot F}{c_{air} \cdot \rho_{air} \cdot V} \cdot \Delta t \cdot [T_{id}(t-1) - T_{od}(t)]}{\frac{1}{K \cdot F} + \frac{1}{c_{air} \cdot \rho_{air} \cdot V} \cdot \Delta t} \quad (13)$$

## 3 Proposed scheduling model

The proposed scheduling model based on the CCP aims to seek the minimum power generation cost of the IES system, and takes into account various operational constraints, especially the probabilistic spinning reserves.

### 3.1 Chance constrained programming

The IES scheduling is a typical stochastic optimization problem, which can be formulated by using the CCP [35]. A general CCP can be expressed as

$$\begin{cases} \min \bar{J} \\ s.t. \ H_j(x) \leq 0 \quad j = 1, 2, ..., NL \\ P_r\{J(x, \xi) \leq \bar{J}\} \geq \beta \\ P_r\{G_k(x, \xi) \leq 0\} \geq \alpha \quad k = 1, 2, ..., NO \end{cases} \quad (14)$$

where $x$ is a decision vector; $J(x,\xi)$ is an objective function; $\xi$ is a vector of random parameters; $G_k(x,\xi)$ is a probabilistic constraint; $P_r\{\cdot\}$ denotes the probability of an event occurring; $\alpha$ and $\beta$ are respectively pre-given confidence levels of the constraint and the objective function; $H$ is a traditional deterministic constraint; $\bar{J}$ is the minimum value of the objective function $J(x,\xi)$ when the function's probability level is not lower than $\beta$; NL and NO denote the total number of the deterministic and probabilistic constraints, respectively.

## 3.2 Formulation of the scheduling model

The proposed scheduling model includes two parts: an objective function and various operational constraints, which will be detailed below.

### 3.2.1 Objective function

In this work, the minimum generation cost of the IES is adopted as the objective function to be optimized. Since the DG outputs in the system are uncontrollable, the spinning reserve costs are considered. Therefore, the used objective function can be expressed as follows:

$$\min F = C_1(P_{it}) + C_2(P_{e,it}) + C_3(P_t^{DH}, P_t^{CH}) \tag{15}$$

$$C_1 = \sum_{t=1}^{T} \sum_{i=1}^{n} [(a_i P_{it}^2 + b_i P_{it} + c_i) + \omega_i R_{it}] \tag{16}$$

$$C_2 = \sum_{t=1}^{T} \sum_{i=1}^{N} \{a_{ir}[P_{e,it} + c_V(P_{h,it}^r + P_{h,it}^c)]^2 + b_{ir}[P_{e,it} + c_V(P_{h,it}^r + P_{h,it}^c)] + c_{ir} + \delta_i R_{e,it}\} \tag{17}$$

$$C_3 = \sum_{t=1}^{T} (g_1(P_t^{DC}) - g_2(P_t^{CH}) + \mu R_t^{BESS}) \tag{18}$$

where $F$ represents the generation cost of the IES; $C_1$ and $C_2$ represent the sum of the fuel costs and the spinning reserve costs of the thermal power units and the CHP units, respectively; $C_3$ represents the sum of the charge and discharge costs and the spinning reserve costs of the BESS; $P_{it}$ is the electric power of thermal power unit $i$ in period $t$; $a_i$, $b_i$, and $c_i$ represent the fuel costs coefficient of thermal power unit $i$, respectively; $P_{e,it}$ is the electric power of the CHP unit $i$ in period $t$; $P_{h,it}^r$ is the total heating power of the CHP unit $i$ with HST in period $t$; $a_{ir}$, $b_{ir}$ and $c_{ir}$ represent the fuel cost coefficient of CHP unit $i$, respectively; $\omega_i$, $\delta_i$, $\mu$, and $R_{it}$, $R_{e,it}$, and $R_t^{BESS}$ represent the factor of spinning reserve costs and spinning reserve of thermal power units, CHP units, and BESS, respectively. It should be noted that the costs of investment and maintenance in this study are not included in the objective function, because these costs are a fixed value and do not affect the scheduling result.

### 3.2.2 Operational constraints

Meeting operational constraints of the IES is a necessary condition for ensuring the IES secure operation. In this section, the operational constraints of the system are depicted in detail.

#### 3.2.2.1 Energy balance constraints

In order to ensure the energy balance of the IES, the demand must be equal to the supply, therefor the electrical balance and heat balance satisfy the following equations [8]:

$$\sum_{i=1}^{n} P_{it} + \sum_{i=1}^{N} P_{e,it} + P_c + P_t^{DC} - P_t^{CH} = P_{lt} + P_{EB,t} \tag{19}$$

$$\sum_{i=1}^{N} (P_{h,it}^r + P_{h,it}^c) + P_{EB,t}^r = P_{lt}^r \tag{20}$$

where $P_c$ is the renewable consumption, which represents the sum of dispatchable wind and PV powers; $P_{lt}$ and $P_{lt}^r$ are the electrical and heating loads of the system in period $t$.

#### 3.2.2.2 Thermal power unit constraints

**Power output constraint:** The output power of the thermal power unit has to be within the following range:

$$P_{i\min} \leq P_{it} \leq P_{i\max} \tag{21}$$

where $P_{i\max}$ and $P_{i\min}$ are the maximum and minimum electric power of thermal power unit $i$, respectively.

**Ramp rate constraint:** When studying system scheduling, the ramp rate limit of each unit is one of the constraints which can't be neglected [23]. The ramp rate constraint is

$$-r_{di} \leq P_{it} - P_{i(t-1)} \leq r_{ui} \tag{22}$$

where $r_{di}$ and $r_{ui}$ are the maximum downward and upward ramp rates of thermal power unit $i$, respectively.

#### 3.2.2.3 Combined heat and power unit constraints

**Output power constraints:** In order to ensure the output power of the CHP unit is within a reasonable range, the electrical output constraint and thermal output constraint can be expressed as follows [9]:

$$P_{e,i\min} \leq P_{e,it} \leq P_{e,i\max} \tag{23}$$

$$0 \leq P_{h,it}^r \leq P_{h,i\max}^r \tag{24}$$

where $P_{e,i\max}$ and $P_{e,i\min}$ are the maximum and minimum electric power of CHP unit $i$, respectively; $P_{h,i\max}^r$ is the upper limit of the heat outputs of the CHP unit.

**Ramp rate constraint:** The ramp rate constraint of each CHP unit must meet the following inequality [23]

$$-r_{di}^r \leq P_{e,it} - P_{e,i(t-1)} \leq r_{ui}^r \tag{25}$$

where $r_{di}^r$ and $r_{ui}^r$ are the maximum downward and upward ramp rates of CHP unit $i$, respectively.

#### 3.2.2.4 Heat storage tank constraints

**Heat charge-discharge rate limit:** The heat charge and discharge power of the HST should obey the following limit:

$$-P_{h,i\min}^{c} \leq P_{h,it}^{c} \leq P_{h,i\max}^{c} \tag{26}$$

where $P_{h,i\max}^{c}$ and $P_{h,i\min}^{c}$ are the maximum and minimum heat storage and heat dissipation of the HST, respectively.

**HST capacity constraints:** The capacity of the HST must be within a reasonable range, in addition, the initial capacity of the HST should be equal to the starting and ending capacities to ensure the initial value of the scheduling in the next cycle is same as the previous one [36]. The above capacity constraints can be expressed as follows:

$$C_{\min} \leq C_t \leq C_{\max} \tag{27}$$

$$C(0)=C(T_{end})=C_{\min} \tag{28}$$

where $C_{\max}$ and $C_{\min}$ are the maximum and minimum heat storage capacity of the HST, respectively; $C(0)$ and $C(T_{end})$ are the starting and ending capacity in one scheduling cycle of the HST, respectively.

#### 3.2.2.5 Battery energy storage system constraints

**Charge and discharge rate constraints:** The charge and discharge power of BESS must be within a reasonable range, which can be expressed as the following inequality [26]:

$$0 \leq P_t^{DC} \leq P_{\max}^{BESS} \tag{29}$$

$$0 \leq P_t^{CH} \leq P_{\max}^{BESS} \tag{30}$$

where $p_{\max}^{BESS}$ is the maximum charge and discharge power.

**BESS capacity constraints:** The capacity constraints of the BESS is the same as the HST, therefor the BESS capacity constraints are [37]

$$S_{\min} \leq S_t \leq S_{\max} \tag{31}$$

$$S(0)=S(T_{end})=S_{\min} \tag{32}$$

where $S_{\max}$ and $S_{\min}$ are the maximum and minimum battery capacities; $S(0)$ and $S(T_{end})$ are the starting and ending capacities in one scheduling cycle, respectively.

#### 3.2.2.6 Electric boiler constraint

For ensuring the safe operation of the EB, its output power should satisfy the following relationship:

$$0 \leq P_{EB,t} \leq P_{EB,\max} \tag{33}$$

where $P_{EB,\max}$ is the maximum power consumption of the EB.

#### 3.2.2.7 Renewable consumption constraint

The amount of renewable energy consumption should be within a reasonable range, which can be expressed as follows:

$$0 \leq P_c \leq E_t \tag{34}$$

where $E_t$ is the expected value of the renewable generation outputs.

#### 3.2.2.8 Spinning reserve constraints

Due to the uncertainty of renewable energies, the system needs spinning reserves to ensure the electric power balance of the system. The spinning reserve provided by thermal power units, CHP units and the BESS obey the following constraints [26]:

$$P_{it} + R_{it} \leq P_{it}^{\max} \tag{35}$$

$$P_{e,it} + R_{e,it} \leq P_{e,it}^{\max} \tag{36}$$

$$R_t^{BESS} \leq \min\{\gamma_{DC}(S(t)-S_{\min})/\Delta t, P_{\max}^{BESS} - P_t^{DC}\} \tag{37}$$

It should be noted here that due to the randomness and volatility of the DG outputs, the spinning reserves of the BESS, the thermal power units and the CHP units must be able to balance the difference between the $E_t$ and the possible random outputs.

In extreme cases, there may be a situation that the DG outputs are 0, so it is necessary to configure a large capacity of spinning reserves. However, in practice, the above extreme cases have a very low probability of occurrence, and the spinning reserve constraint can be described in the form of CCP [31].

$$P_r\{\sum_{i=1}^{n}R_{it} + \sum_{i=1}^{N}R_{e,it} + R_t^{BESS} \geq E_t - P_t^W - P_t^{PV}\} \geq \alpha \tag{38}$$

## 4 Model solving

This section firstly generates a probabilistic sequence based on the probability distribution of DGs outputs according to the SOT [38], and then the chance constraint in the model is transformed into its deterministic equivalence class, which has the linear structure of MILP, and finally solved by the CPLEX solver.

### 4.1 Probability serialization representation of distributed generation outputs

A sequence can be defined as a series of values at non-negative integer points on the number axis. For a sequence $a(i_a)$ with length $N_a$ ($i_a = 0,1,...,N_a,...$), it is named a probabilistic sequence if the following condition is satisfied:

$$\sum_{i_a=0}^{N_a} a(i_a) = 1, \quad a(i_a) \geq 0 \tag{39}$$

The WT outputs and PV outputs are random variables since the probability distribution is known, the probabilistic sequence $a(i_{at})$ and $b(i_{bt})$ can be obtained respectively. The formation of a probabilistic sequence can be considered as a special discretization of the probability distribution of continuity. The length $N_{at}$ of the PV outputs probabilistic sequence $a(i_{at})$ is [31]

$$N_{at} = \lceil P_{PV\max t} / q \rceil \tag{40}$$

where $\lceil \cdot \rceil$ is the ceiling function; $P_{PV\max t}$ is the maximum possible outputs value of the PV in period $t$; $q$ is the discrete step size; After discretization, there are a total of $N_{at}+1$ states, in which the output of the state $m_a$ is $m_a q (0 \leq m_a \leq N_{at})$, and the corresponding probability is $a(m_a)$. Therefore, the PV output and its corresponding probabilistic sequence $a(i_{at})$ are shown in Table 1.

Table 1. PV output and its corresponding probabilistic sequence

| Power (MW) | 0 | $q$ | ... | $m_a q$ | ... | $N_{at} q$ |
|---|---|---|---|---|---|---|
| Probability | $a(0)$ | $a(1)$ | ... | $a(m_a)$ | ... | $a(N_{at})$ |

The probabilistic sequence $a(i_{at})$ is calculated by using the probability density function $f_p(P_{PV})$ of PV outputs.

$$a(i_{at}) = \begin{cases} \int_0^{q/2} f_p(P_{PV}) dP_{PV}, & i_{at} = 0 \\ \int_{i_{at}q-q/2}^{i_{at}q+q/2} f_p(P_{PV}) dP_{PV}, & i_{at} > 0, i_{at} \neq N_{at} \\ \int_{i_{at}q-q/2}^{i_{at}q} f_p(P_{PV}) dP_{PV}, & i_{at} = N_{at} \end{cases} \tag{41}$$

Similarity, the probabilistic sequence $b(i_{bt})$ corresponding to the WT outputs in period $t$ is handled in the same way as the probabilistic sequence corresponding to the PV outputs.

### 4.2 Conversion of chance constraint to deterministic equivalence

There are two random variables, i.e. $P_t^{PV}$ and $P_t^W$, with different distributions in Eq. (38). The prerequisite for the conversion from the chance constraint in Eq. (38) into its deterministic equivalence class is that the distribution of random variable $Z = P_t^{PV} + P_t^W$ must be available. The distribution of variable Z is as follows [26]:

$$F_Z(z) = \int_{-\infty}^{z} [\int_{-\infty}^{\infty} f_w(z-y) f_p(y) dy] dz \tag{42}$$

where $f_w(\cdot)$ is the probability density of WT outputs, $f_p(\cdot)$ is the probability density of PV outputs. Only after obtaining $F_Z^{-1}(z)$ through the inverse transformation can the deterministic transformation of chance constraints be implemented, but $F_Z^{-1}(z)$ is difficult to be obtained due to the complex forms of $f_w(\cdot)$ and $f_p(\cdot)$. Through utilizing the SOT to handle probability distribution of the variable Z, the transformation from chance constraint to its deterministic class can be successfully achieved.

The probabilistic sequence $c(i_{ct})$ corresponding to the DG outputs in period $t$ can be obtained by [21]

$$c(i_{ct}) = \sum_{i_{at}+i_{bt}=i_{ct}} a(i_{at}) b(i_{bt}), \quad i_{ct} = 0, 1, ..., N_{at} + N_{bt} \tag{43}$$

The DG output sequence that takes $q$ as a discrete step and $N_{ct}$ ($N_{ct} = N_{at} + N_{bt}$) as the length is shown in Table 2. Each output corresponds to a probability, which together constitutes a probabilistic sequence of DG output $c(i_{ct})$.

Table 2. The DG output and its corresponding probabilistic sequence

| Power (MW) | 0 | $q$ | $2q$ | ... | $(N_{ct}-1)q$ | $N_{ct} q$ |
|---|---|---|---|---|---|---|
| Probability | $c(0)$ | $c(1)$ | $c(2)$ | ... | $c(N_{ct}-1)$ | $c(N_{ct})$ |

To easily handle Eq. (38), a new class of 0-1 variable $z_{m_{ct}}$ is defined as [26]

$$z_{m_{ct}} = \begin{cases} 1, & \sum_{i=1}^{n} R_{it} + \sum_{i=1}^{N} R_{e,it} + R_t^{BESS} \geq E_t - m_{ct} q, \; m_{ct} = 0, 1, ..., N_{at} + N_{bt} \\ 0, & otherwise \end{cases} \tag{44}$$

Eq. (44) shows that in period $t$, if the total spinning reserve is not less than the difference between the expected value $E_t$ and the DG output $m_{ct}q$, $z_{m_{ct}}$ is taken as 1; otherwise, it is taken as 0.

According to Table 2, Eq. (38) can be simplified as

$$\sum_{m_{ct}=0}^{N_{at}+N_{bt}} z_{m_{ct}} c(m_{ct}) \geq \alpha \tag{45}$$

Taking into account that $z_{m_{ct}}$ cannot be directly solved by using MILP, Eq. (44) can be transformed into

$$(\sum_{i=1}^{n} R_{it} + \sum_{i=1}^{N} R_{e,it} + R_t^{BESS} + m_{ct}q - E_t)/L \leq Z_{m_{ct}} \leq 1 + (\sum_{i=1}^{n} R_{it} + \sum_{i=1}^{N} R_{e,it} + R_t^{BESS} + m_{ct}q - E_t)/L, m_{ct} = 0,1,...,N_{at}+N_{bt} \tag{46}$$

where $L$ is a large positive number. Given $\sum_{i=1}^{n} R_{it} + \sum_{i=1}^{N} R_{e,it} + R_t^{BESS} \geq E_t - m_{ct}q$, Eq. (46) is equivalent to $\lambda \leq z_{m_{ct}} \leq 1 + \lambda$ ($\lambda$ is a very small positive number). Since $z_{m_{ct}}$ is a 0-1 variable, it can only be equal to 1; otherwise, it is equal to 0. In this way, the original CCP-based scheduling model is reformulated as a readily solvable MILP structure.

### 4.3 Solution process

The flowchart of the proposed method is shown in Fig. 4. Specifically speaking, the solution process mainly includes the following steps:

Step 1: Build the structure model of the IES.
Step 2: Establish an IES optimal scheduling model based on CCP.
Step 3: Discretize the probability density function of the DG outputs and generate the corresponding probabilistic sequences.
Step 4: Based on the SOT, obtain the expected value of DG outputs in each time period.
Step 5: Transform the probabilistic spinning reserve constraint into its deterministic equivalence class.
Step 6: According to the membership function obtained by the fuzzy of heating load, the indoor thermal comfort temperature range is determined.
Step 7: The heating load demand of buildings is calculated by the thermal transient equilibrium equation.
Step 8: Reformulate the original CCP-based scheduling model into the MILP form.
Step 9: Input parameters of the IES.
Step 10: Use CPLEX optimization software to solve the model and obtain the optimal solution.
Step 11: Check whether the solution exists. If found, terminate the solution process; otherwise, update the confidence level and go to step 9.
Step 12: Output the IES optimal scheduling scheme.

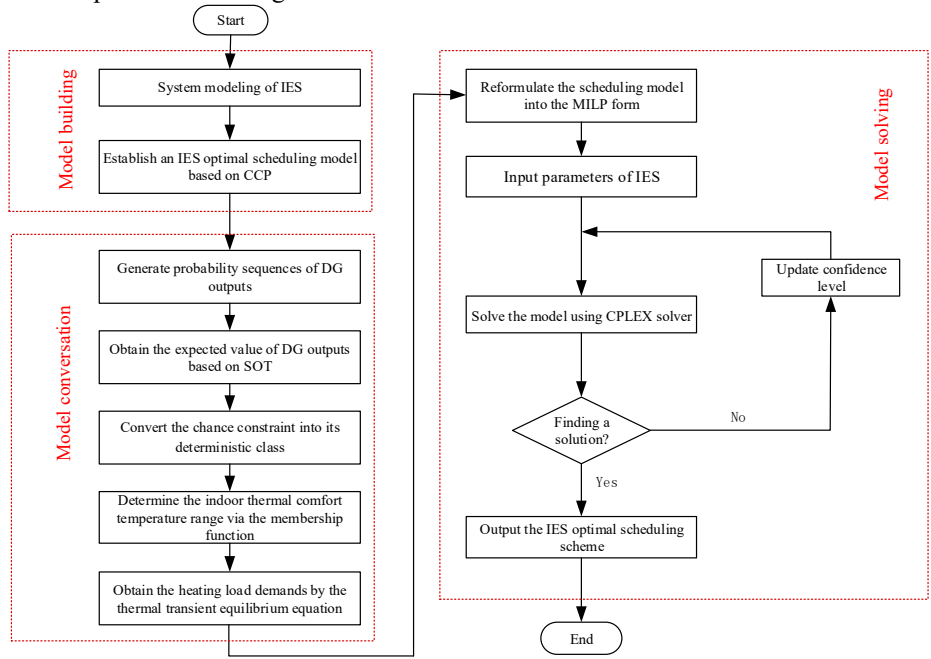

Fig. 4. Flowchart of the proposed method

## 5 Case studies

In this paper, the proposed scheduling approach is tested on a modified IEEE 30-bus system, where the electrical subsystem is modeled at the distribution network scale, and the heating subsystem is modeled at the district level. All procedures are developed by using MATLAB R2016b with the commercial optimization solver CPLEX. All simulations are performed on a PC platform with 8 GB RAM and 2 Intel Core dual-core CPUs (2.4 GHz).

## 5.1 Introduction of the test system

As shown in Fig. 5, the generators 1 and 2 are replaced by two CHP units, the WT and PV are respectively connected to buses 6 and 17 in the test system. In addition, the system is equipped with an EB, a BESS and two HSTs with the same parameters. The specific parameters of the thermal and CHP units are shown in Tables 3 and 4.

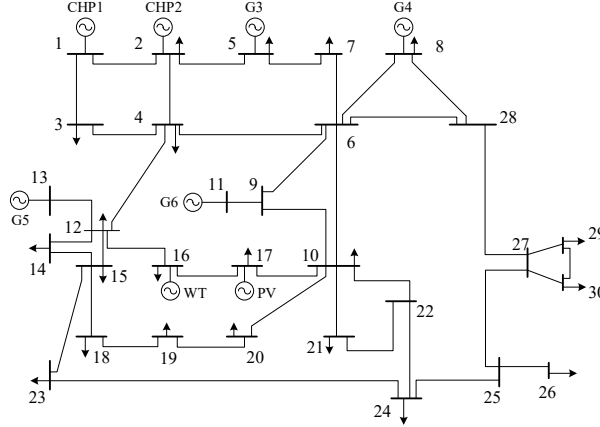

Fig. 5. One-line diagram of the modified IEEE 30-bus system

Table 3. Main parameters of thermal power units

| $P_{max}$ (MW) | $P_{min}$ (MW) | $r_{ui}$ (MW/h) | $r_{di}$ (MW/h) | $a_i$ ($/MW²) | $b_i$ ($/MW) | $c_i$ ($) | $\omega_i$ ($/MW) |
|---|---|---|---|---|---|---|---|
| 50 | 25 | 25 | 25 | 0.012 | 17.82 | 10.150 | 13.7 |
| 35 | 10 | 18 | 18 | 0.069 | 26.24 | 31.670 | 13.2 |
| 30 | 10 | 15 | 15 | 0.028 | 37.69 | 17.940 | 13.2 |
| 40 | 12 | 20 | 20 | 0.010 | 12.88 | 6.778 | 14.2 |

Table 4. Main parameters of CHP units

| $P_{e,max}$ (MW) | $P_{e,min}$ (MW) | $P_{h\,max}^{r}$ (MW) | $r_{ui}^{r}$ (MW/h) | $r_{di}^{r}$ (MW/h) | $a_{ir}$ [$/(MWh)⁻¹] | $b_{ir}$ [$/(MWh)⁻¹] | $c_{ir}$ ($/h⁻¹) | $C_V$ | $C_m$ | $\delta_i$ ($/MW) |
|---|---|---|---|---|---|---|---|---|---|---|
| 200 | 100 | 250 | 50 | 50 | 0.0044 | 13.29 | 39 | 0.15 | 0.75 | 16.2 |
| 200 | 100 | 250 | 50 | 50 | 0.0044 | 13.29 | 39 | 0.15 | 0.75 | 16.2 |

Other parameters used in this study are set as follows. The building parameters are $K=0.5\text{W}\cdot\text{m}^{-2}\cdot°\text{C}$, $F=2.3\times10^7\text{m}^2$, $V=5\times10^7\text{m}^3$, $c_{air}=1.007\text{kJ}\cdot\text{kg}^{-1}\cdot°\text{C}$, $\rho_{air}=1.2\text{kg}\cdot\text{m}^{-3}$; the WT parameters are $v_{in}=3\text{m/s}$, $v_r=15\text{m/s}$, $v_{out}=25\text{m/s}$, $p_r=60\text{kW}$; the PV parameters are $\eta=0.093$, $A_c=1300\text{m}^2$, $p_{max}^{PV}=120\text{kW}$; the BESS parameters: $S_{max}=160\text{MW}\cdot\text{h}$, $S_{min}=32\text{MW}\cdot\text{h}$, $p_{max}^{BESS}=40\text{MW}$, $\gamma_{CH}=\gamma_{DC}=0.9$ [26], charge and discharge costs are $100\$/\text{MW}\cdot\text{h}$ and $150\$/\text{MW}\cdot\text{h}$ respectively, the spinning reserve cost is $20\$/\text{MW}\cdot\text{h}$; the HST parameters are $C_{max}=240\text{MW}\cdot\text{h}$, $C_{min}=40\text{MW}\cdot\text{h}$, $P_{hi\,max}^{c}=P_{hi\,min}^{c}=50\text{MW}$; the EB parameters are $P_{EB,max}^{r}=30\text{MW}$, $\eta_{ah}=0.95$; the scheduling cycle $T=24$; $\Delta t=1\text{h}$. In this study, the predicted electrical load powers in a scheduling cycle are shown in Table 5.

Table 5. Predicted values of electric load power in different periods

| Time period | Power (MW) | Time period | Power (MW) | Time period | Power (MW) | Time period | Power (MW) |
|---|---|---|---|---|---|---|---|
| 1 | 400 | 7 | 400 | 13 | 600 | 19 | 440 |
| 2 | 390 | 8 | 430 | 14 | 590 | 20 | 400 |
| 3 | 380 | 9 | 470 | 15 | 580 | 21 | 390 |
| 4 | 410 | 10 | 540 | 16 | 570 | 22 | 380 |
| 5 | 390 | 11 | 580 | 17 | 560 | 23 | 390 |
| 6 | 400 | 12 | 592 | 18 | 520 | 24 | 400 |

In this paper, the losses of the BESS, the HST, and the lines are neglected. To verify the validity of the model, the following modes are used to simulate and compare the simulation. Among the modes, modes 1-3 simultaneously consider the uncertainty of renewable generations and the thermal inertia of buildings; while modes 4 and 5 only take into account the thermal inertia of buildings and renewable uncertainties, respectively, and mode 6 does not consider both aspects.

Mode 1: The traditional method of "heat-set" electro-thermal coupling scheduling. The BESS, the HST, and the EB are not involved in the dispatch, and the CHP units alone bear the heating load demands.

Mode 2: The BESS participates in the system power supply, the HST and the EB don't participate in the heating, and the CHP units still bear the system heating load demands alone.

Mode 3: Each CHP unit is equipped with a HST, and both the BESS and the EB participate in the system optimal scheduling.

Mode 4: The system is equipped with the above three kinds of auxiliary equipments, but the uncertainty of renewable generations is not considered. That is mean the DGs are modeled as deterministic form, the spinning reserves are set as 20% of the system's net load which is the difference between the original electric load and the expected value of DG outputs.

Mode 5: The system is equipped with three kinds of auxiliary equipments, but the thermal inertia of buildings is not considered.

Mode 6: The system is equipped with three kinds of auxiliary equipments, but neither the uncertainty of renewable generations nor the thermal inertia of buildings is considered. The spinning reserves are set as same as mode 4.

Table 6 shows the details of the six different operating modes.

Table 6. The six different operating modes

| Operation modes | BESS | HST | HB | Uncertainties of DGs | Thermal inertia of buildings |
|---|---|---|---|---|---|
| Mode 1 | × | × | × | √ | √ |
| Mode 2 | √ | × | × | √ | √ |
| **Mode 3** | **√** | **√** | **√** | **√** | **√** |
| Mode 4 | √ | √ | √ | × | √ |
| Mode 5 | √ | √ | √ | √ | × |
| Mode 6 | √ | √ | √ | × | × |

## 5.2 Results and analysis

In order to properly analyze the influence of different factors, such as three kinds of auxiliary equipments, DG uncertainties and the thermal inertia, on the operation of the IES, the above-mentioned different modes are adopted for subsequent comparative analysis. The specific analysis process is as follows.

### 5.2.1 Analysis of load demands and distributed generation expected outputs

In this study, the winter solstice day with low outdoor temperature is selected as a typical day. For ease of analysis, the indoor temperature is maintained at 20 °C. The heating load demand varies with the outdoor temperature, which is shown in Fig. 6.

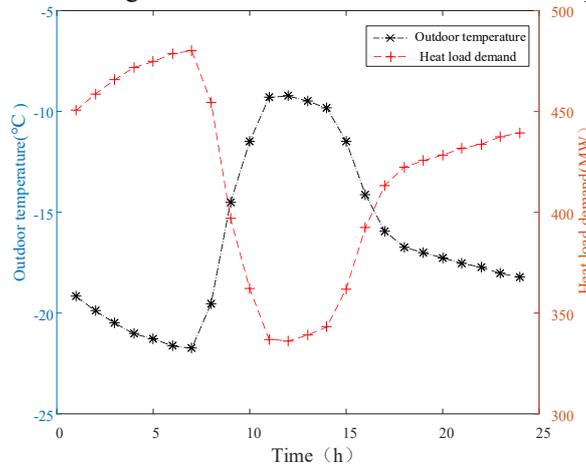

Fig. 6. Outdoor temperature and heating load demand

It can be seen from Fig. 6 that if the indoor temperature does not change, the lower the outdoor temperature, the greater the heating load demands; the higher the outdoor temperature, the less the heating load demands. The reason for this is that the lower the outdoor temperature, the greater the difference between the indoor and outdoor temperature, the greater the heat dissipation of the building. In order to maintain the indoor temperature during this period, the building needs to absorb more heat, whereas the building absorbs less heat. Therefore, it can be concluded that the heating load model can effectively depict the heating load demands that meet the thermal comfort of the human body.

The expected value of the joint outputs of DGs and the daily changes of the heat and electrical loads in a scheduling cycle are shown in Fig. 7.

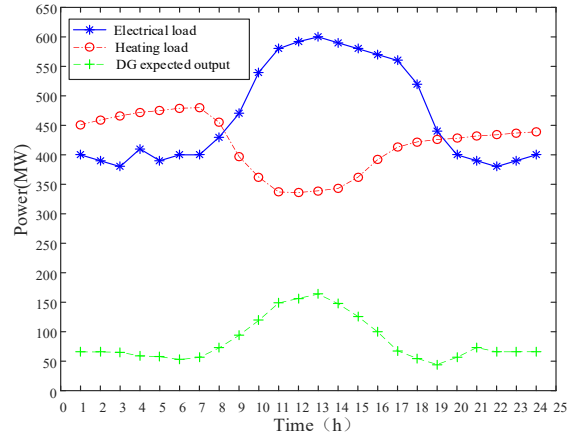

Fig. 7. Expected value of DG joint outputs and daily changes of heat and electrical loads

From Fig. 7, it can be found that the following phenomena: 1) the expected value of DG outputs accounts for about 20% of the total electrical loads, and the output is low at night because there is no PV power generation; 2) the electricity loads and the heating loads are complementary in time. When the electricity loads increase, the heating loads decrease, and when the electricity loads decrease, the heating loads increase. This is because the electricity demand during the day is more than that of heating, and the heat demand at night is more than that of electricity.

### 5.2.2 Analysis of confidence levels

In order to choose a suitable confidence level of probabilistic spinning reserve constraints, the spinning reserve capacities and the generation costs under different confidence levels are analyzed with the results shown in Figs. 8 and 9.

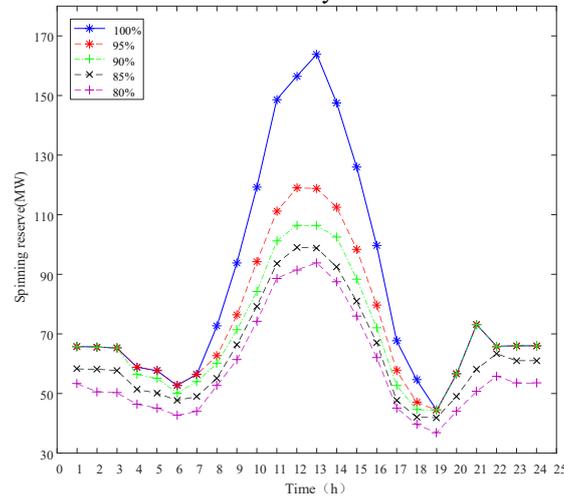

Fig. 8. Reserve capacities with different confidence levels

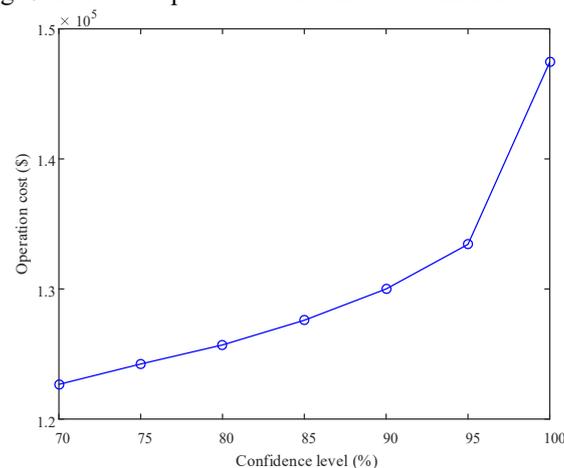

Fig. 9. Generation costs with different confidence levels

It is known from Figs. 8 and 9 that, when the confidence level increases, the more spinning reserve are required and the higher reliability of the system; However, the higher the confidence level, the higher the generation costs of the system; especially after the confidence level reaches 95%, if the confidence level continues to increase, the generation costs increase dramatically. Therefore, choosing an appropriate confidence level is very important to trade off economy and reliability. Based on the above considerations, the relevant modes in this paper all adopt the confidence level of 95%.

### 5.2.3 Economic analysis

For the purpose of reasonably evaluating the operational economy of the system under the six modes, this study compares and analyzes the generation costs in different modes with the results shown in Table 7.

Table 7. Costs in different modes

| Operation modes | costs ($) |
|---|---|
| Mode 1 | 195160.57 |
| Mode 2 | 135239.17 |
| **Mode 3** | **133431.02** |
| Mode 4 | 135278.15 |
| Mode 5 | 134243.46 |
| Mode 6 | 135647.13 |

It can be seen from Table 7 that the generation cost in mode 1 is the highest, the generation cost in mode 3 is the lowest, and the generation costs in other modes lie halfway in between. In particular, the total cost in mode 3 reaches the minimum value of 133431.02$, which is decreased by 61,729.55$ compared with the conventional mode 1, and is decreased by 1808.15$ compared with mode 2. It can be seen that the combined operation with HST, BESS and EB can achieve the best economy. In addition, the generation costs of modes 4, 5 and 6 are respectively 1847.13$, 812.44$ and 2216.11$ greater than that of mode 3, it can be concluded that considering the uncertainty of DGs and the thermal inertia of a building simultaneously can minimize the generation costs and ensure the operational economy of the system.

### 5.2.4 Analysis of the auxiliary equipments

In order to understand the influence of BESS, HST and EB on renewable consumption and system optimization scheduling, this paper separately analyzes the renewable energy consumption and the outputs of the three representative auxiliary equipments in a scheduling cycle in modes 1-3.

#### 5.2.4.1 Impact of auxiliary equipments on renewable energy consumptions

Renewable energy consumptions in different modes are shown in Fig.10. The details of renewable consumption of mode 3 are shown in Fig. 11.

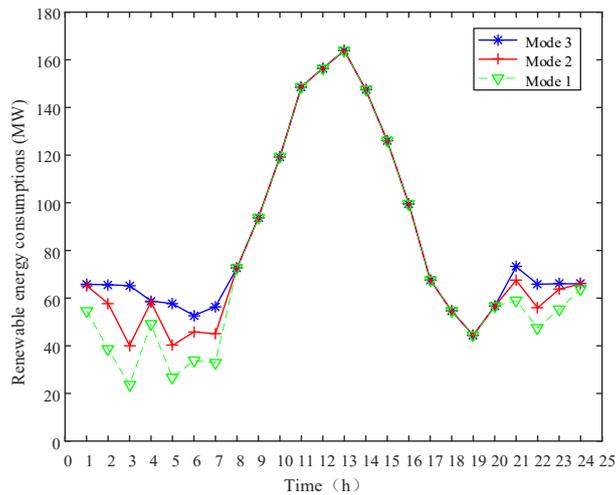

Fig. 10. Renewable energy consumptions in different modes

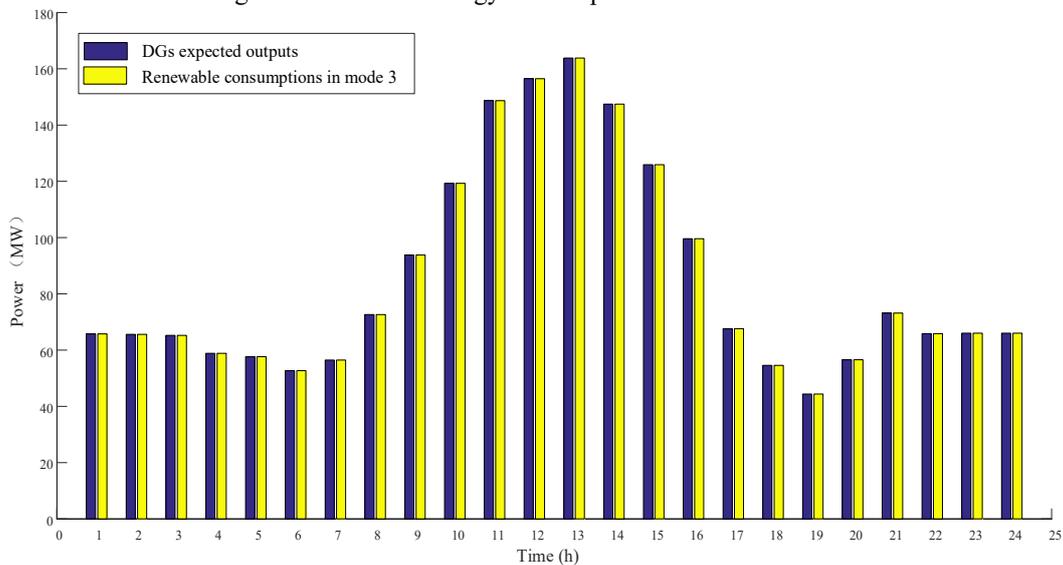

Fig. 11. Renewable energy consumptions in mode 3

It can be observed from Fig. 10 that in mode 1, a large amount of renewable curtailment occurs during the night; in mode 2, the amount of renewable consumption has increased; and the amount of renewable consumption reaches the maximum value in mode 3. Figs. 10 and 11 show that renewable energy resources are completely consumed during the periods 8:00-20:00 periods in the three modes since the electricity load is significantly higher than the heating load in these periods and thereby DG outputs ease the thermal-electric constraints of CHP units; while the electricity load is lower than the heating load during the nighttime hours so that there are a lot of curtailments of renewable energy resources. Compared with mode 1, a BESS is equipped in mode 2, so the flexibility of the system is raised; compared with mode 2, two HSTs and an EB are equipped in mode 3, so the flexibility of the system is further enhanced, thereby reducing renewable curtailments. Furthermore, as can be easily seen from Fig. 11, the renewable curtailments are completely absorbed in mode 3. This shows that the proposed method can solve the proposed scheduling model effectively. By utilizing energy storage devices such as HSTs, EBs, and BESS, the flexibility of the system operation is improved, and the level of renewable consumption is promoted, it provides an effective way to solve the problem of renewable curtailments.

#### 5.2.4.2 Analysis of energy storage

The coordinated operation of energy storage with multiple energy forms can improve the operational flexibility of the system by thermal-electric decoupling and realizes the complementary advantages of multi-energy storage, which contributes to the safe and economic operation of the system. As a tie linking the thermal and electric sub-systems, the energy storage can improve the rigid connection between the two sub-systems and achieve decoupling of them in the time and space dimensions. The outputs of heat storage and battery energy storage are analyzed as follows.

#### 5.2.4.2.1 Outputs of the battery energy storage system

The charge and discharge powers of the BESS in modes 2 and 3 are shown in Fig. 12.

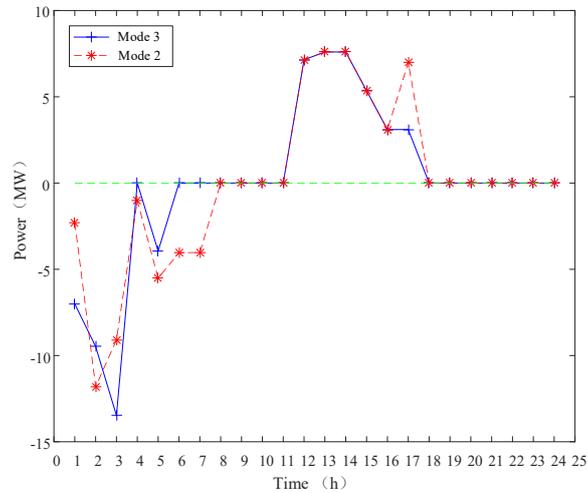

Fig. 12. Charge and discharge power of BESS

It can be seen from Fig. 12 that the BESS charged during the period from 0:00 to 6:00, and discharged during the period 11:00-18:00, at other times, the capacity of the BESS remains unchanged. This is because when the electricity demands increase during the day, the BESS can be equivalent to the power supply and discharge to the user; when the electricity demands decrease in the evening, the BESS can be equivalent to the load and charged by the system. By appropriately shifting the electric load, the BESS can actively participate in the overall scheduling of the system, achieving the peak-load shifting effects.

#### 5.2.4.2.2 Outputs of the heat storage tanks

The heat storage and release powers of the HSTs in mode 3 are shown in Fig. 13.

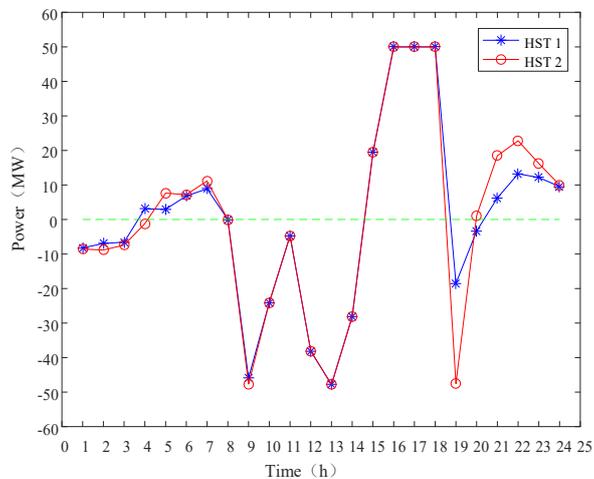

Fig. 13. Heat storage and release power of HSTs under mode 3

It can be seen from Fig. 13 that HSTs in mode 3 have different heat storage and release power conditions at different time periods. Specifically, during the period 8:00-14:00, HSTs perform heat storage, during the periods of 4: 00-8:00, 15: 00-18:00 and 20: 00-24:00, HSTs are used for heating. The reason is that when the DGs outputs are large and the power loads are small, the CHP units use the heat in HSTs to reduce outputs of the units (including the heat outputs and the electric outputs), thereby the space for the DGs power consumption is increased. When the electric loads are large and the heating load demands are small, the CHP units increase their outputs. In addition to meeting the heating loads, the excess heat is stored in HSTs. It should be noted that during the period 1:00-4:00, the reason why HSTs store heat is that the initial capacities of HSTs are assumed to be the minimum capacities allowed, so in the initial period, the HSTs only heat storage without heat release.

In the case of unchanged the heating load on the user side, the HST can adjust the heating load of the CHP units. Especially, reducing the heating power of the CHP units at night can expand the adjustable range of the power outputs of the units, thereby mitigating the constraints of the heating load on their power outputs to a certain extent. As long as HST can store enough heat in the period of lower heating load, the CHP units can reduce the power output to the minimum level as much as possible during the electrical load valley periods, which can effectively improve the peak regulation capacity.

#### 5.2.4.3 Outputs of the electric boiler

In mode 3, the heat outputs of the CHP units, the HST and EB are shown in Fig. 14.

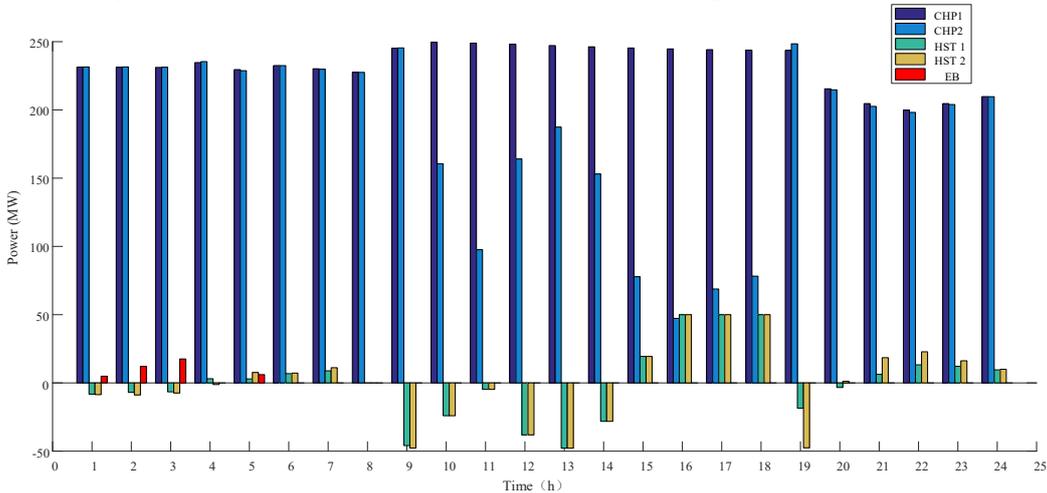

Fig. 14. The heat outputs of each device in mode 3

It can be seen from Fig. 14 that in the period from 1:00 to 5:00, since the initial capacity of the HST is assumed to be the minimum capacity allowed, all the heat released by the CHP units and HSTs cannot meet the heat demands of the system, in order to maintain heat balance, the EB performs heat release. In other time periods, the heating load has already met the heat balance, so the EB is not involved in system regulation.

#### 5.2.4.4 Complementary and coordinated operation of auxiliary equipments

In order to further analyze the impacts of complementary and coordinated operation of multiple auxiliary equipments on the system operation, the outputs of multiple auxiliary equipments in mode 3 are shown in Fig. 15.

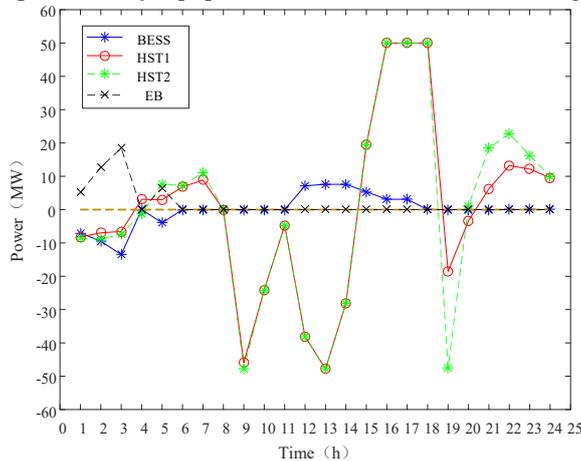

Fig. 15. The outputs of each auxiliary equipment in mode 3

It can be observed from Fig. 15 that multiple auxiliary equipments are in the complementary operation during the periods 1:00-6:00 and 11:00-15:00; and in the coordinated operation during the period 15:00-18:00; while during other periods, only the HSTs operate. When the electricity demands decrease and heating demands increase in the evening, the BESS is charged by the system, and the CHP units decrease electric outputs due to the reducement of their heat outputs, thereby improving the space for the renewable energy consumption, while the HSTs and EB release heats to meet heating demands; at noon, the heating demands decrease and the electricity demands increase, so the BESS discharges to the user and the HSTs perform heat storage. When there are more electricity and heating demands, the BESS and HSTs simultaneously release energy. Base on these facts, it can be safely concluded that the complementary and coordinated operation of multiple auxiliary equipments not only can improve the system operational flexibility, but also can enhance the reliability of the energy supply. The HSTs and BESS have different charging and discharging periods such that there is always a certain energy reserve in each period during system operation, which contributes to strengthen the reliability of the energy supply.

#### 5.2.5 Analysis of the characteristics of renewable generations and buildings

The uncertainties of renewable generations and the thermal inertia in buildings are two different characteristics that can affect the optimized operation of the system. Therefore, this paper analyzes the above two characteristics separately.

**5.2.5.1 Analysis of renewable generation uncertainties**

In order to prove the necessity of considering the uncertainty of renewable generations, the spinning reserves provided in modes 3 and 4 are compared in Fig. 16.

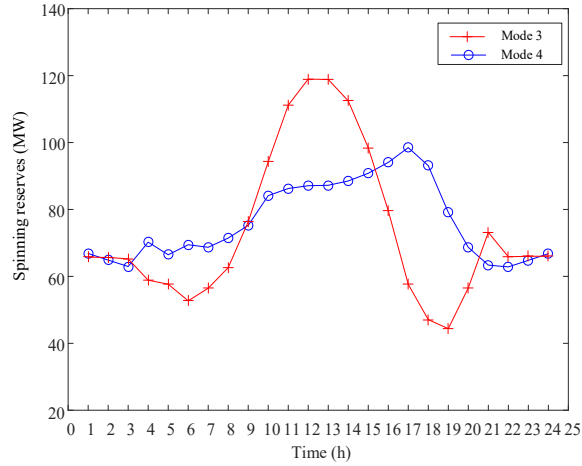

Fig. 16. Reserve capacities in different situations

From Fig. 16, it can be found that the spinning reserves provided by the system are more flexible when considering the uncertainty of DGs. The form of spinning reserves is the probabilistic constraint related to a risk of a constraint violation, rather than traditional deterministic constraints. During the period 9:00-15:00, since the DG outputs are large the system needs to provide more spinning reserves to fill the difference between the actual and predicted value of renewable generations due to their intermittency and uncertainty. However, when the uncertainty of DGs is not considered, the spinning reserves provided by the system are obviously insufficient, which will cause the safety of the system to decrease. During the period 3:00-9:00 and 15:00-20:00, the DG outputs are low, when not considering the uncertainty of DGs, the spinning reserves are overestimated, resulting in unnecessary spinning reserves are wasted. Therefore, the uncertainty model of DGs is better than the deterministic one.

Note that, the above analysis is based on the consideration of thermal inertia. The results of the comparison with modes 5 and 6 are the same as the above results.

**5.2.5.2 Analysis of the thermal inertia in buildings**

In order to analyze the impact of the thermal inertia in buildings on system operation, renewable energy consumptions in modes 3, 5 and modes 4, 6 are compared respectively, which is shown in Fig. 17 and Fig. 18.

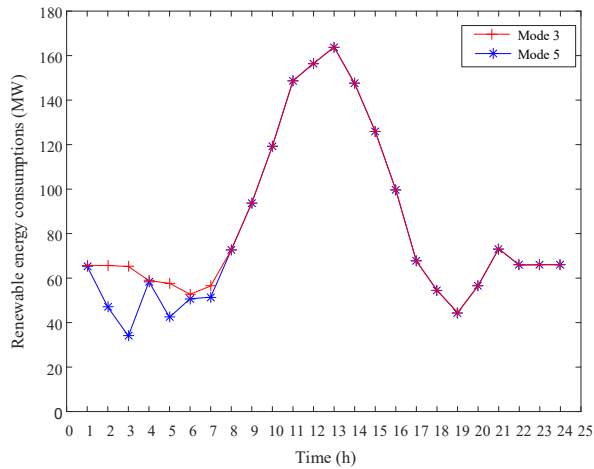

Fig. 17. Renewable energy consumptions in different situations

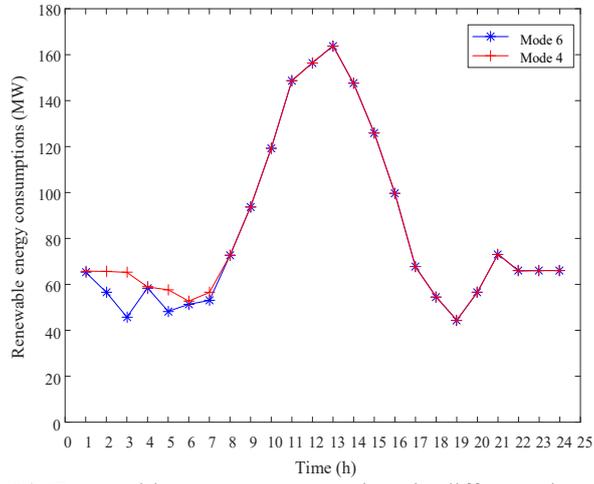

Fig. 18. Renewable energy consumptions in different situations

It can be observed from Figs. 17 and 18, from 1:00 to 7:00, the amount of renewable absorption when considering the thermal inertia is more than that when not considering the thermal inertia in buildings. This is because the envelope structure of a building can store a certain amount of heat which means that the building can be regarded as a heat storage unit with a certain heat storage capacity. At night, the electric load demand is relatively small, so a large amount of renewable curtailment occurs, and due to the thermal inertia in buildings, the units' electric outputs can be reduced to a certain extent, thereby the space for the DGs power consumption is increased. Therefore, it can be concluded that considering the thermal inertia of the building can improve the operating flexibility of the system and promote renewable energy consumption. So far, mode 3 is the best operation mode for system scheduling in this study.

### 5.2.6 Analysis of the electric output of each unit

In order to understand intuitively the change of each unit outputs under the optimal operation schedule with time, this paper analyzes the electric outputs of each unit under mode 3 with the results shown in Fig. 19.

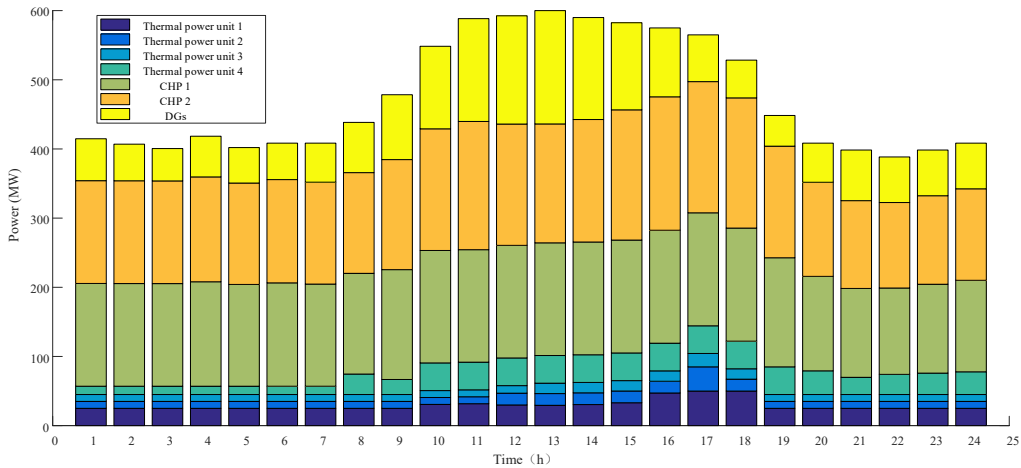

Fig. 19. The output of each unit in mode 3

As shown in Fig. 19, since the fuel cost of the CHP units is lower than the fuel cost of the thermal power units, the CHP units provide more electric outputs during the peak load periods. Among the thermal power units, since the fuel costs of the thermal power units 1 and 4 are lower than those of the thermal power units 2 and 3, the electric power outputs of the units 1 and 4 are preferentially increased.

### 5.2.7 Analysis of calculation time

To examine the computational efficiency of the proposed method, the calculation times of the different modes with and without consideration of renewable uncertainties are shown in Table 8.

Table 8. Calculation times of different modes

| Operation modes | Calculation times (s) | | | |
| --- | --- | --- | --- | --- |
| | Confidence levels (Consideration of renewable uncertainties) | | | Without consideration of renewable uncertainties |
| | 95% | 90% | 85% | |
| Mode 1 | 6.360 | 6.589 | 6.485 | — |
| Mode 2 | 6.548 | 6.636 | 6.683 | — |
| Mode 3 | 6.472 | 6.633 | 6.554 | — |
| Mode 5 | 6.559 | 6.465 | 6.621 | — |
| Mode 4 | — | — | — | 6.483 |
| Mode 6 | — | — | — | 6.527 |

As shown in Table. 8, the calculation times of the proposed approach at different modes are about 6.5 seconds which meets the real-time requirements of the IES operation. It can be expected that in practical applications, the calculating time can be further reduced the proposed approach by using more advanced computer hardware and an optimized code implemented in a low-level programming language like C and FORTRAN. Therefore, it can be seen that the computational efficiency of the proposed method meets the real-time requirements for handling optimization scheduling of IES in real-world applications.

## 6  Conclusion

As insufficient flexibility in system operation resulted from traditional "heat-set" operating modes of CHP units restricts renewable energy consumption, how to improve the operational flexibility through thermal-electric decoupling has nowadays become an urgent and challenging problem. In this regard, a novel optimal scheduling model based on chance-constrained programming is proposed to seek the minimum generation cost for a small-scale integrated energy system. In this model, taking renewable uncertainty into account, the probabilistic spinning reserves are provided by thermal power units, CHP units and BESS; and a heating load model is built by considering thermal comfort in buildings. To resolve the model, a SOT-based solution approach is developed. Based on the simulation results, the following conclusions can be drawn:

(1) The presented IES scheduling model manages to improve the operational flexibility of the system with uncertain renewable generations by comprehensively leveraging thermal inertia of buildings and auxiliary equipments, which provides a fundamental way to promote renewable energy consumption.

(2) The developed solution approach is able to solve the proposed model accurately and quickly. By transforming the chance constraint into its deterministic equivalence class, the original chance-constrained programming-based scheduling model is capable of being converting into a readily solvable MILP formulation, and then solved by utilizing the efficient CPLEX solver.

(3) As thermal-electric decoupling measures, the auxiliary equipments such as HST, EB and BESS have different effects on improving the system operational flexibility. Furthermore, for a given IES, auxiliary equipments should be properly selected according to the specific electricity and heating demands of the system to achieve optimal performance. In addition, the complementary and coordinated operation of different auxiliary equipments is not only able to improve the system operational flexibility, but also able to enhance the reliability of the energy supply.

In future work, an integrated scheduling model of renewable generation and demand response will be introduced into integrated energy systems. In addition, this study has not considered the capacity losses of energy storages, time delay and transmission losses in IES operation, while a more realistic scenario shall take into account this effect during optimization. Besides, with the popularization of electric vehicles, it is an interesting topic to investigate the integration of vehicle-to-grid in the integrated energy system.

### Acknowledgments

This work is partly supported by the "13th Five-Year" Scientific Research Planning Project of Jilin Province Department of Education under Grant No. JJKH20200113KJ, the China Scholarship Council (CSC) under Grant No. 201608220144 and the National Natural Science Foundation of China under Grant No. 51677023.